# Thermal transport in two-dimensional C$_3$N/C$_2$N superlattices: A molecular dynamics approach

Leila Razzaghi[a], Farhad Khoeini[a*], Ali Rajabpour[b], Maryam Khalkhali[a]

[a] Department of Physics, University of Zanjan, Zanjan, 45195-313, Iran

[b] Advanced Simulation and Computing Lab. (ASCL), Mechanical Engineering Department, Imam Khomeini International University, Qazvin, 34148–96818, Iran

**ABSTRACT**: Nanostructured superlattices have been the focus of many researchers due to their physical and manipulatable properties. They aim to find promising materials for new electronic and thermoelectric devices. In the present study, we investigate the thermal conductivity of two-dimensional (2D) C$_3$N/ C$_2$N superlattices using non-equilibrium molecular dynamics. We analyze the dependence of thermal conductivity on the total length, temperature, and the temperature difference between thermal baths for the superlattices. The minimum thermal conductivity and the phonon mean free path at a superlattice period of 5.2 nm are 23.2W/m.K and 24.7 nm, respectively. Our results show that at a specific total length, as the period increases, the number of interfaces decreases, thus the total thermal resistance decreases, and the effective thermal conductivity of the system increases. We found that at long lengths ($L_x$ >80 nm), the high-frequency and low-wavelength phonons are scattered throughout the interfaces, while at short lengths, there is a wave interference that reduces the thermal conductivity. The combination of these two effects, i.e., the wave interference and the interface scattering, is the reason for the existence of a minimum thermal conductivity in superlattices.

1. Introduction

Thermal conductivity is one of the basic materials properties, which is a crucial parameter in designing of micro/nanoelectronic devices. After the discovery of graphene [1], its properties, such as unique thermal and mechanical properties [2-5], and the ability to engineer the structural and electronic properties, motivated researchers to study new two-dimensional materials, including carbon-nitride nanostructures [6-9]. In the 1990s, many attempts were made by researchers to synthesize carbon-nitride structures, in various ways, which is a new class of two-dimensional materials. Carbon-nitride 2D nanostructures have a non-zero electronic energy bandgap, which makes them outstanding candidates for future applications of the next-generation electronic devices [10-12]. In recent years, the heat transfer property of carbon-nitride nanostructures with impurities, interfaces, and defects at the atomic scale has been studied [13-15]. A system containing periodic potential is called a superlattice, namely, a periodic structure composed of two or more materials [16, 17].

In the past few decades, nanostructured superlattices, due to their physical properties and the possibility of engineering structural properties in electronic devices, have attracted the attention of researchers [18-20].

A superlattice can have different properties than each of its components. The thermal conductivity of superlattices has many applications in thermoelectric and optoelectronic devices, and is suitable for improving the energy converters of these devices with great importance [21, 22]. The thermal conductivity of superlattices has been investigated in many experimental and theoretical works. Among these recent experimental works, we can mention the GaAs/AlAs superlattices [23, 24], perovskite oxide superlattices [25], and the TiN/(Al, Sc)N superlattices [26].

Molecular dynamics simulations have been performed on superlattices of graphene-hBN [27], germanium nanowire [28], graphene-nitrogenated holey graphene [29], silicone nanowires [30, 31], and many other types of superlattices [18, 32-35]. The results of all these studies show that superlattices can have lower thermal conductivity than any of their components. In general, the scattering of phonons in altered structures reduces the heat flow. In superlattices, the scattering of phonons reduces the thermal conductivity due to the reflection and transmission of the thermal vibrations at the interface. Interface conditions are very important in controlling heat transfer. In smooth contacts, the effect of wave interference is predominant, but in rough contacts, phonons are more scattered. Another factor that affects the thermal conductivity of a superlattice, is the period length. Competition between the period length and the mean free path of phonons can lead to a decrease in the thermal conductivity of a superlattice [36]. Recent studies, performed by molecular dynamics simulations, have shown that the thermal conductivity of graphene-hBN superlattices reduces by up to 98% compared to pure graphene, depending on the period of the superlattice [27].

In this work, we investigate the heat transport in superlattices of the $C_3N/C_2N$ via non-equilibrium molecular dynamics simulations, as shown in Fig. 1. We examine the dependence of the thermal conductivity on the total length, period length, mean temperature of the system, and the temperature difference applied to thermal baths. By observing the non-uniform behavior of the thermal conductivity versus the superlattice period, we determine a minimum value for the thermal conductivity of the superlattice. Moreover, non-equilibrium molecular dynamics simulations are employed to calculate the interfacial thermal resistance at the $C_3N/C_2N$ interface.

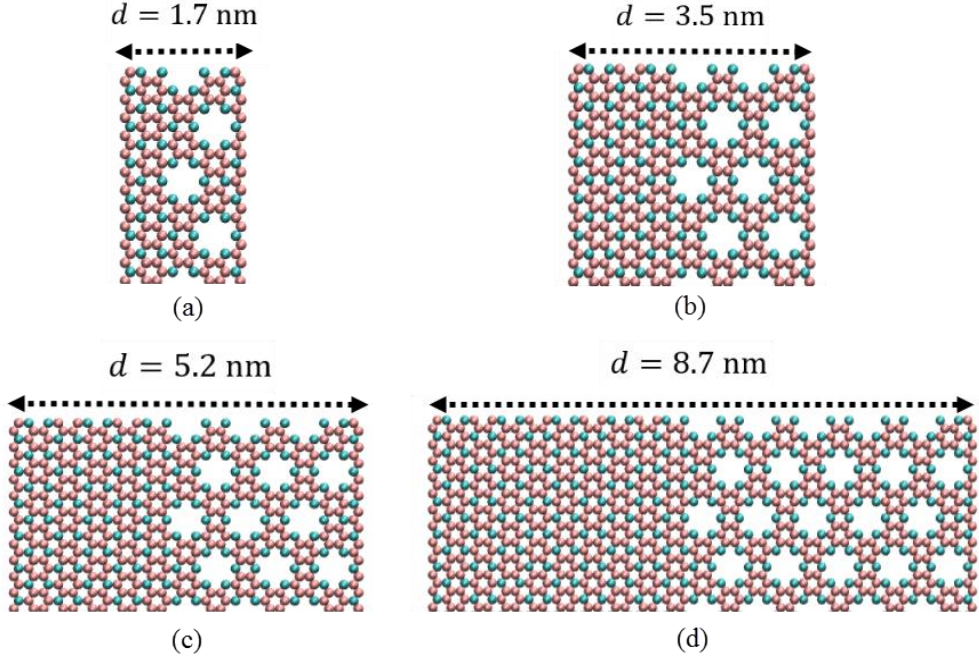

Fig. 1. The unit cell of $C_3N/C_2N$ structures with increasing the length of superlattice period, $d$. Carbon and nitrogen atoms are rendered in pink and cyan, respectively.

## 2. Simulation method

In this study, we have employed molecular dynamics (MD) simulations with the large-scale atomic/molecular massively parallel simulator (LAMMPS) package, to compute the thermal transport properties in $C_3N/C_2N$ superlattices [37]. We have considered the Tersoff potential function, based on the optimized coefficients by Lindsay and Brodio [38, 39]. It is worth noting that Newton's equations of motion were integrated via the velocity Verlet algorithm [40] with a time step of 1 fs. Also, the periodic boundary condition is employed in the X and Y directions, and a free boundary condition is set in the Z direction. In this study, we calculate the effective thermal conductivity for the $C_3N/C_2N$ superlattice. To this end, energy minimization is first performed to adjust the coordinates of the atoms in the system. In the next step, the structure is relaxed at room temperature, 300 K, for 1 ns under the NVE ensemble and Langevin thermostat. Then, to create a temperature gradient and measuring the heat flow, the system is divided into slabs with a thickness of 1 nm. To prevent the system from rotating about the X direction, the final slabs are fixed during the simulation. Next to these fixed slabs, the hot and cold regions are respectively, set to temperatures of $T + \Delta T$ (320 K) and $T - \Delta T$ (280 K), via the Nose-Hoover thermostat [41], under the NVT ensemble. The rest of the slabs are considered in the NVE ensemble. In the following stage, the heat flow ($q_x$) is calculated using the slope of the energy lines when the system reaches the non-equilibrium steady state. By measuring the energy variations of the thermostats versus the simulation time, and considering the Fourier's law, the thermal conductivity of the system in the X direction is obtained as follows:

$$q_x = -\kappa A \frac{dT}{dx} \tag{1}$$

where $\kappa$ is the thermal conductivity of the structure, and $A$ is the cross-section area. Besides, $q_x$ is the heat current, and $\frac{dT}{dx}$ is the temperature gradient along the X direction. It is worth noting that, the system is simulated for 6 ns after thermal relaxation. Molecular dynamics setup for calculating the interfacial resistance of the $C_3N/C_2N$ nanostructure is shown in Fig. 2. The interfacial thermal resistance or Kapitza resistance (R) is computed as follows:

$$R = \frac{A * \Delta T_{int}}{q_x} \tag{2}$$

where $\Delta T_{int}$ is the temperature jump at the interface and can be obtained from the temperature profile along the sample length [42]. The thickness of carbon-nitride nanostructures ($C_3N$ and $C_2N$) is assumed to be 3.4 Å.

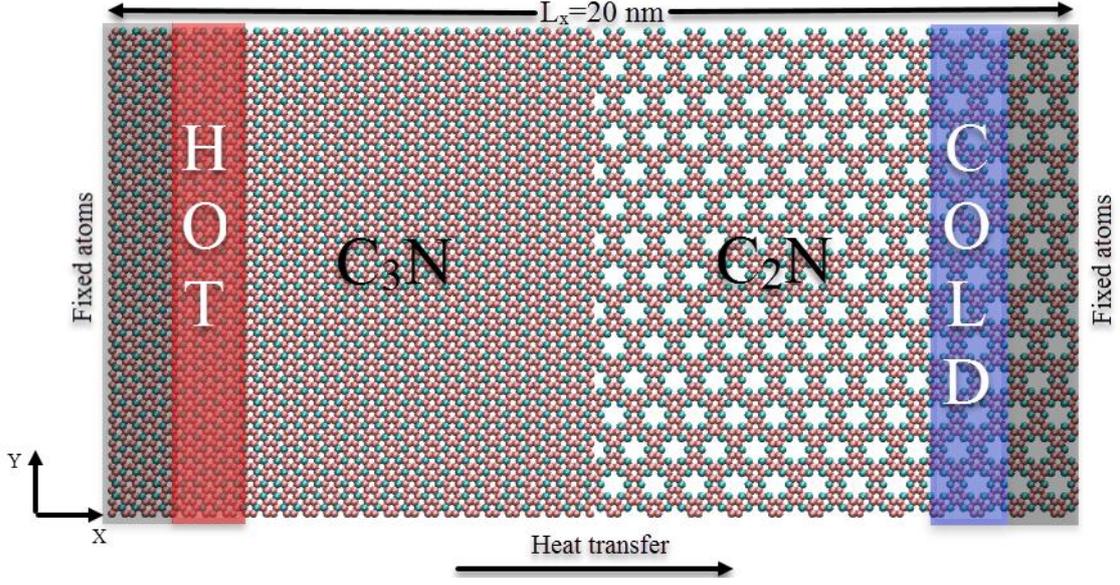

Fig. 2. Molecular dynamics setup for evaluating the interfacial resistance of the $C_3N/C_2N$ system. Carbon and nitrogen atoms are rendered in pink and cyan, respectively.

## 3. Results and discussion

### 3.1. Length dependence of the thermal conductivity

The thermal conductivity of a system with a finite length of $L_x$ can be related to its bulk value as shown in the following equation [43]:

$$\frac{1}{\kappa(L_x)} = \frac{1}{\kappa_\infty}\left(1 + \frac{\Lambda_{ph}}{L_x}\right) \qquad (3)$$

where $\kappa_\infty$ indicates the bulk thermal conductivity, and $\Lambda_{ph}$ is the effective phonon mean free path (MFP). Using the data obtained from the simulation, we can calculate the infinite thermal conductivity and effective phonon mean free path for each period of the superlattice. This thermal conductivity is the same as the effective thermal conductivity for each period of the superlattice. Molecular dynamics results show that the thermal conductivity increases as the total length for each superlattice period increases. Figure 3 shows the length dependency of the thermal conductivity of the superlattice for four period lengths from 1.7 nm to 8.7 nm. As shown in figure 3, three types of regimes can be considered for the $C_3N/C_2N$, as also proposed by Perierra et al. [27] for graphene/h-BN superlattices. First, the ballistic regime, which is valid for small superlattice lengths (< 40 nm). In this regime, the system length is shorter than the phonon mean free path of the system. Second, at lengths longer than 80 nm, the dependence of the thermal conductivity on the length is weak, and a diffusive regime is observed. Finally, there is a regime between these two ballistic and diffusive regimes, called the transmission regime for medium lengths. In this regime, the thermal conductivity is slightly dependent on the length of the system, which is comparable to the phonon mean free paths. In figure 3, the uncertainties were calculated through five simulations with different initial conditions.

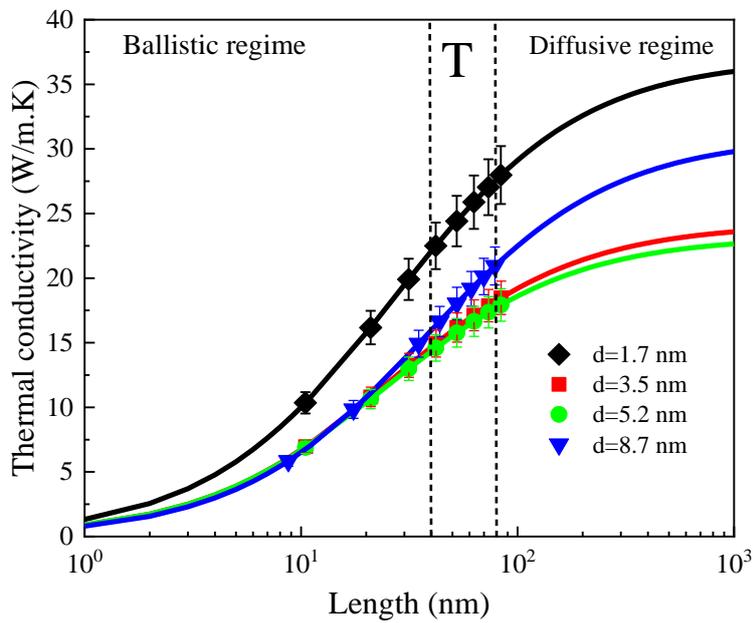

Fig.3. Length dependency of the thermal conductivity for various period lengths. The dashed lines indicate the ballistic transport regime, the transmission regime (T), and the diffusive regime. The error bars for all system sizes are calculated based on five different simulation runs.

### 3.2. Effect of superlattice period

Figure 4 shows the bulk thermal conductivity $\kappa_\infty$, and effective phonon mean free path of the $C_3N/C_2N$ superlattice versus the superlattice period, $d$, at a temperature of 300 K.

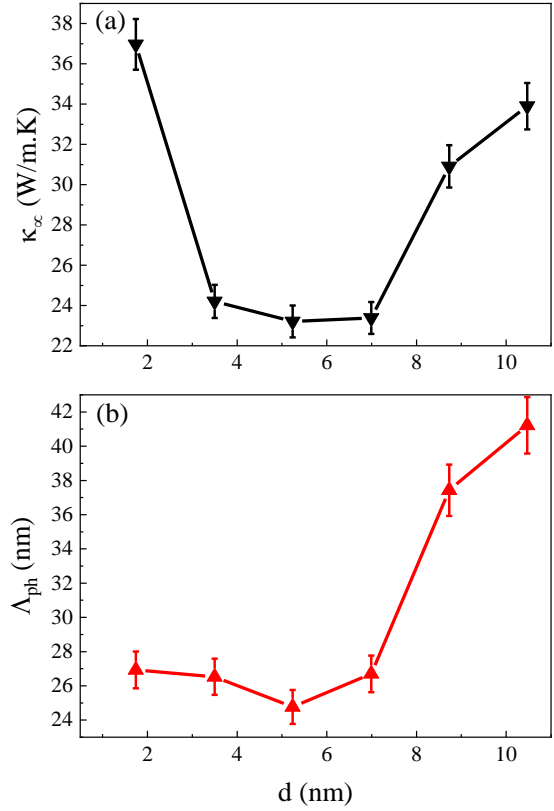

Fig. 4. Dependency of (a) the bulk thermal conductivity, and (b) the effective phonon MFP versus the superlattice period. The error bars for all system sizes are calculated based on five different simulation runs.

By analyzing the results, we observe that the overall thermal conductivity of the superlattice decreases, compared to the thermal conductivity of the $C_3N$ and $C_2N$. The thermal conductivity of the superlattice is reduced up to ≈ 97%, compared with the thermal conductivity of the $C_3N$. Also, it is reduced up to ≈ 60%, compared with the thermal conductivity of the $C_2N$ [44]. As it is shown in figure 4 (a), there is a nonlinear dependence of $\kappa_\infty$ on the $d$. Initially, with increasing the d, the thermal conductivity increases, then at $d$= 5.2 nm, it reaches its minimum value, about $\kappa_\infty$ = 23.2 W/m.K, and then increases again. The obtained results are in agreement with the previous studies by Perriera et. al. [27]. Moreover, in previous researches, a minimum for the thermal conductivity versus the period length/thickness has been observed for different types of superlattices [32, 35, 45-47]. However, the value of minimum thermal conductivity could be numerically different, due to the difference in the length and width of the supercell. To investigate the thermal rectification of the $C_3N/C_2N$ system with $L_x$=20 nm at T = 300 K, and heat baths temperature difference of 40 K, we impose hot and cold (cold and hot) heat baths to the left and right (right and left). Figure 5 shows the temperature distribution for the 2D $C_3N/C_2N$ system. The temperature distribution of the structure is linear in slabs further away from the thermal baths, while, it is nonlinear in the region adjacent to the thermal baths, which is due to the strong scattering of phonons. Also, temperature drops at the interface are observed in the temperature profiles of the structure. The magnitude of these temperature drops are about 10.9 K in (+) direction and about 7.6 K in (-) direction. The temperature drops are due to the scattering of vibrational carriers at the interface of the two structures. The values of $R^+$ and $R^-$ for the $C_3N/C_2N$ interface are about $2.89 \times 10^{-9}$ m²K/W and about $2.10 \times 10^{-9}$ m²K/W, respectively.

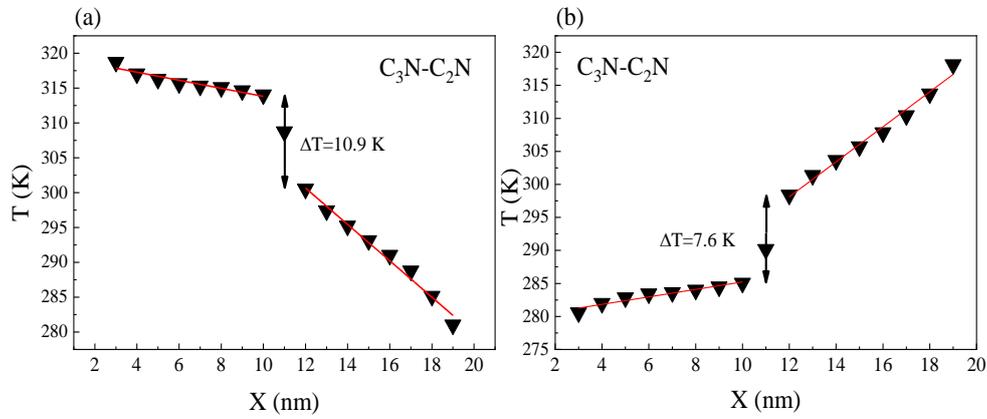

Fig. 5. The steady-state of 1D temperature profiles for heat flux in (a) (+) direction and (b) (-) direction with $L_x$=20 nm at T = 300 K, and the heat baths temperature difference of 40 K.

The corresponding values $R^+$ and $R^-$ indicate that the interfacial thermal resistance also depends on the heat flux direction. We define thermal rectification as:

$$TR = \frac{R^+ - R^-}{R^-} \times 100\% = 37.6\% \tag{4}$$

To better understand the differences in the interfacial thermal resistance value for the $C_3N/C_2N$ interface, the phonon power spectral density of the structure is illustrated in Fig. 6.

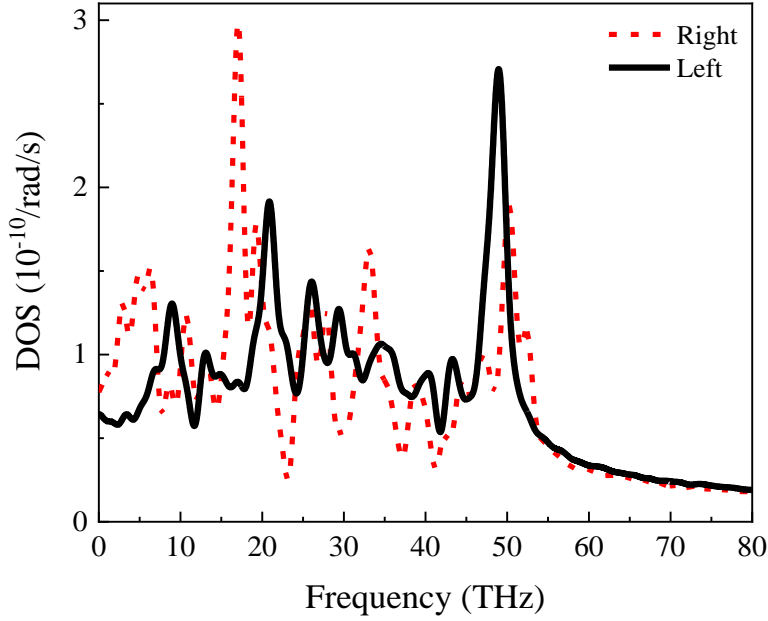

Fig. 6. Phonon power spectral density of states on two sides of the $C_3N/C_2N$ interface at T = 300 K.

The phonon density of the states is obtained by computing the Fourier transform of the autocorrelation function of the velocity of atoms corresponding to the $C_3N/C_2N$ interface, as follows [48, 49]:

$$P(\omega) = \sum_j \frac{m_j}{k_B T} \int_0^\infty e^{-i\omega t} < \boldsymbol{v}_j(t).\boldsymbol{v}_j(0) > dt, \qquad (5)$$

where $m_j$, $\boldsymbol{v}_j$, and $\omega$ are the mass, the velocity of atom j, and the angular frequency, respectively. As shown in figure 6, there is a strong mismatch between the left and right spectra. This asymmetry of the phonon spectrum justifies the scattering of phonons at the interface, resulting in a thermal resistance at the interface.

As the mean temperature of the system increases, two things happen. First, energetic phonons are created in the system since the mobility of the phonons is high. So, the system has higher frequency phonons. If there are many high-energy phonons in a system, more energy is transferred from the hot zone to the cold one. This factor causes more energy transfer and, ultimately, increases the thermal conductivity (see Fig.7). It is concluded that growing the temperature increases the overall thermal conductivity of the system. Also, with increasing temperature, the kapitza resistance decreases, and the thermal conductivity increases. Second, growing the temperature increases the phonon-phonon interaction, and the scattering between them, which leads to a decrease in the thermal conductivity [50]. If one of these two factors overcomes the other, it dominates the thermal behavior of the system. Figure 7 shows the thermal conductivity of two superlattices with different lengths versus the temperature. It is found that in both cases the thermal conductivity decreases by increasing the temperature, and then increases at higher temperatures, due to the dominance of the high-energy phonons mechanism. It is also revealed that the superlattice with $L_x = 83.9\ nm$ has higher thermal conductivity values at all temperatures than a superlattice with $L_x = 52.4\ nm$. In both cases, the period length is 5.2 nm.

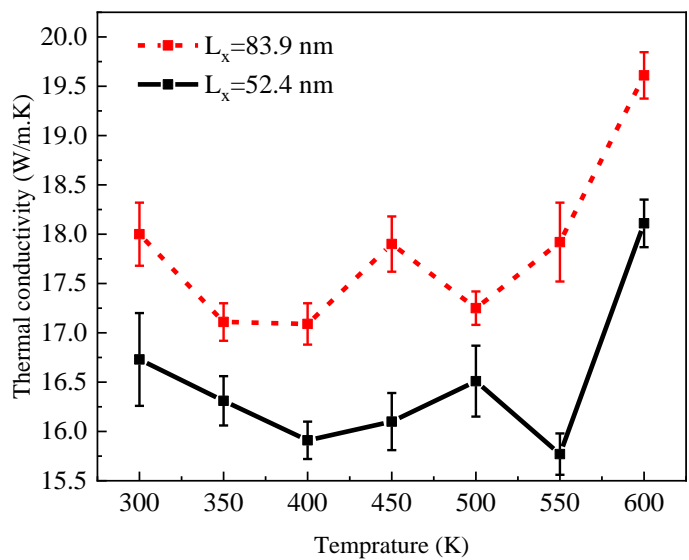

Fig.7. Thermal conductivity in terms of temperature for two superlattice lengths of $L_x = 52.4\ nm$ and $L_x = 83.9\ nm$. The error bars are calculated based on five different simulation runs.

Figure 8 shows the thermal conductivity in terms of the temperature difference between the thermal baths at the mean temperature of 300 K. Since, there are different energetic phonons in various temperature differences between the thermal baths, and, consequently, different phonon-phonon scatterings, so these factors affect the heat transfer and the thermal conductivity of the system, as shown in figure 8.

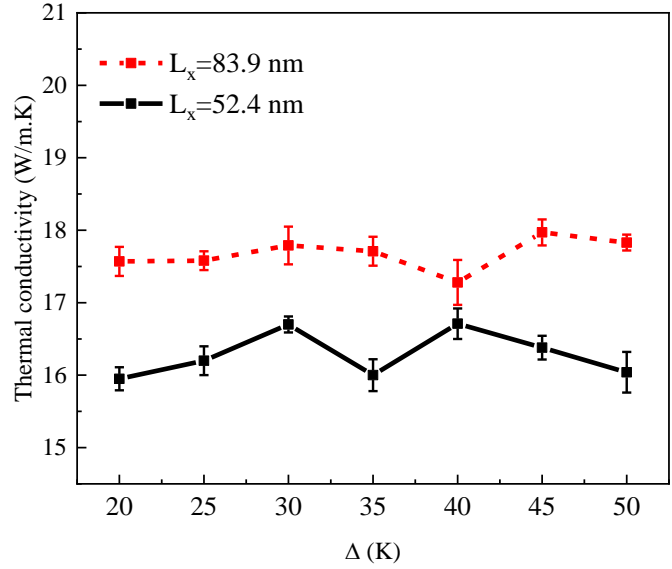

Fig. 8. Thermal conductivity in terms of the temperature difference, $\delta$, between thermal baths for two superlattice lengths of $L_x = 52.4\ nm$ and $L_x = 83.9\ nm$ at the mean temperature of 300 K. The error bars are calculated based on five different simulation runs.

## 4. Conclusions

In this study, we investigated the dependence of the thermal conductivity on the total length, temperature and temperature difference between thermal baths for the $C_3N/C_2N$ superlattice. Extensive non-equilibrium molecular dynamics simulations were conducted to explore the thermal conductivity of the superlattice utilizing the Tersoff potential function. Our results indicate that the minimum thermal conductivity and effective phonon mean free path at room temperature were estimated to be almost 23.2 W/m.K and 24.7 nm, respectively, by utilizing the Tersoff interatomic potential. This minimum occurs, because although with decreasing the period length, the thermal conductivity decreases, because of the increase in the number of interfaces, but in the short period length, due to the interference of heat waves, the interface loses its effect, and the thermal conductivity increases. Moreover, it was found that the thermal conductivity of the 2D $C_3N/C_2N$ superlattices at infinite length is reduced up to 97%, compared with the thermal conductivity of the $C_3N$, and also, it is reduced up to 60%, compared with the thermal conductivity of the $C_2N$. The values of Kapitza resistance at the $C_3N/C_2N$ interface was also computed to be about $2.89 \times 10^{-9}$ $m^2$ K/W. The results of this study can be useful in the application of carbon-nitride superlattices in the thermal design of nanoelectronic components.

**Declaration of Competing Interest**

The authors declared that there is no conflict of interest.

Farhad Khoeini (khoeini@znu.ac.ir)